\documentclass[11pt]{article}
\usepackage[margin=1in]{geometry}
\usepackage{graphicx} 
\usepackage[hidelinks]{hyperref}
\usepackage{amsmath} 
\usepackage[numbers]{natbib}
\usepackage{setspace}
\usepackage{booktabs}
\usepackage{multirow}
\usepackage[table]{xcolor}
\usepackage{lscape}
\usepackage{colortbl}
\usepackage{array}
\usepackage{longtable} 
\usepackage{float}
\usepackage{amsfonts}

\title{\textbf{Refining and Robust Backtesting of \textit{A Century of Profitable Industry Trends}}}

\author{
    Alessandro Massaad\\
    am6535
    \and
    Rene Moawad\\
    rm3892
    \and
    Oumaima Nijad Fares\\
    onf2102
    \and
    Sahaphon Vairungroj\\
    sv2744
    \\
    \\
    Columbia University
    \\
    \\
    Under the mentorship of Prof. Naftali Cohen
}

\date{December 2024}

\begin{document}

\maketitle

\begin{abstract}
We revisit the long-only trend-following strategy presented in \textit{A Century of Profitable Industry Trends} by Zarattini and Antonacci \cite{zarattini2024century}, which achieved exceptional historical performance with an 18.2\% annualized return and a Sharpe Ratio of 1.39. While the results outperformed benchmarks, practical implementation raises concerns about robustness and evolving market conditions. This study explores modifications addressing reliance on T-bills, alternative fallback allocations, and industry exclusions. Despite attempts to enhance adaptability through momentum signals, parameter optimization, and Walk-Forward Analysis, results reveal persistent challenges. The results highlight challenges in adapting historical strategies to modern markets and offer insights for future trend-following frameworks. 
\end{abstract}

\section{Introduction}
Momentum strategies are categorized into time-series momentum, which relies on an asset's own historical returns, and cross-sectional momentum, which compares returns across different assets. Industry momentum, focusing on sector-based trends, has demonstrated significant profitability, often surpassing individual stock momentum over intermediate periods. \textit{A Century of Profitable Industry Trend} by Zarattini and Antonacci \cite{zarattini2024century} examines a long-only trend-following strategy across 48 U.S. industry portfolios from 1926 to 2024, reporting an average annual return of 18.2\% with a Sharpe Ratio of 1.39, outperforming the U.S. equity market's 9.7\% return and 0.63 Sharpe Ratio.  However, the study's methodology presents challenges in practical implementation and raises questions about robustness, particularly concerning the impact of market conditions on the strategy's performance.

Two primary challenges in the original strategy merit further investigation. First, The reliance on T-bills as the default risk-free allocation poses practical challenges for real-world application, as large-scale portfolios rarely achieve full T-bill exposure. Second, the strategy’s calibration over the entire historical dataset raises potential concerns about overfitting, which may affect predictive power in dynamic market conditions. The lack of robust validation further undermines its applicability to out-of-sample scenarios.

This project is a modification of an existing industry momentum strategy, focusing on addressing its practical limitations and improving robustness through different approaches. Rather than relying solely on T-bills for risk mitigation, we explore alternative fallback allocations that maintain market exposure while controlling downside risk. We implement Walk-Forward Analysis to dynamically retrain and validate the strategy over time, aiming to reduce overfitting and improve adaptability to evolving market conditions. We optimize key parameters—including rolling periods and volatility targets—and refine industry selection methods to improve risk-adjusted returns. While this project explores iterative enhancements to address practical limitations—such as reliance on T-bills, overfitting, and evolving market conditions—results highlight the persistent challenges of achieving robust, generalizable performance. This study aims to provide insights into these challenges, offering a foundation for future research into adaptive trend-following strategies.

\section{Data}

\subsection{Data Sources}
We use two publicly available sources from Kenneth French's Data Library. Both datasets were accessed as compressed CSV archives, and their detailed references are provided in the bibliography:
\begin{itemize}
\item \textbf{48 Industry Portfolios}: The industry portfolio return database is constructed using data sourced from Kenneth French’s online data library. Each industry portfolio is rebalanced annually based on market capitalization, ensuring representative and dynamic weighting over time. The database spans the period from 2000 to 2024 and provides daily total returns for 48 industries. At the start of each year t, NYSE, AMEX, and NASDAQ stocks are assigned to industry portfolios based on their four-digit SIC codes from the previous year (t-1), ensuring the portfolios are defined on a point-in-time basis. In cases where SIC codes are unavailable, alternative sources are employed to maintain dataset consistency. Stocks are removed from an industry portfolio immediately upon their CRSP delisting date \cite{french_industry_portfolios}.
\item \textbf{Fama-French Research Factors}: The dataset includes daily market risk factors, such as the combined market return and risk-free Treasury bill (T-bill) rates \cite{french_research_factors}.
\end{itemize}

\subsection{Data Collection Period}
To ensure rigorous testing and validation, we split the data into four distinct sets:
\begin{itemize}
\item In-sample (2000–2015): Used to train and optimize the strategy, ensuring it captures historical patterns and trends.
\item In-sample Validation (2016–2020): Used for validation within the same sample but over a later period, to test whether the model generalizes to more recent trends.
\item Validation (2021–2022): A distinct dataset to assess generalizability across varying market conditions, ensuring the strategy's adaptability. 
\item Out-of-sample (2023--2024): The most recent data, used to evaluate real-world applicability under new conditions, as referenced by \texttt{pt1\_data}, \texttt{pt2\_data}, \texttt{pt3\_data}, in the fallback strategy implementation \cite{mia2}. 

\end{itemize}

By incorporating these distinct data sets, we ensure that the strategy undergoes robust evaluation, balancing academic rigor with practical applicability in dynamic market environments. 

\subsection{Point-In-Time Compliance}
Point-in-time (PIT) compliance is rigorously maintained throughout the analysis to ensure that no future information is used in decision-making. The methodology relies on explicitly lagging signals and weights, which aligns all data with the timing of real-world observations. 
\begin{itemize}
    \item \textbf{Lagging Signals:} Signals are shifted by one period using \texttt{.shift(1)} to ensure they represent only past data. For instance, long band signals are lagged using \texttt{long\_band\_shifted = long\_band.shift(1)}, and similarly, short band signal are lagged with \texttt{short\_band\_shifted = short\_band.shift(1)}. These shifted values are then used to generate trading signals without incorporating future information.
    \item \textbf{Lagging Portfolio Weights:} Portfolio weights are lagged using \texttt{.shift(1)} to ensure that allocations are based on the previous period's data. For instance, portfolio weights are shifted using \texttt{port[f'weight\_{i+1}'].shift(1).fillna(0)}, ensuring that returns are computed based on lagged weights.
    \item \textbf{Returns Calculation:} To maintain PIT compliance in cumulative return calculations, any missing values in percentage changes are backfilled using \texttt{fillna(0)} \\ (e.g., \texttt{ret = aum['AUM'].pct\_change(fill\_method=None).fillna(0)}). This ensures continuity without introducing forward-looking bias. 
\end{itemize}.

\subsection{Data Preprocessing}
\paragraph{Handling Missing Values:}
\begin{itemize}
    \item Missing values in raw data are converted to NaN using \texttt{pd.to\_numeric(errors='coerce')}.
    \item Returns below -99\% are treated as invalid and replaced with NaN.
    \item Missing values during cumulative calculations are filled with zeroes to ensure continuity.
\end{itemize}
Rows with missing values in key columns such as \texttt{mkt\_ret} or \texttt{tbill\_ret} are excluded before analysis.

\section{Original Strategy Overview }

The original strategy aims to construct a portfolio capable of capturing long-term upward trends while minimizing downside risk. It employs a systematic rules-based approach consisting of three components: entry criteria, position sizing, and exit criteria. In this section, we give a quick summary of the original strategy. For a detailed description, see \cite{zarattini2024century}.

\subsection{Summary of Key Components}

\subsubsection{Entry Criteria}
The strategy utilizes Keltner Channels and Donchian Channels to generate entry signals. These indicators are designed to identify trends while allowing for price oscillations within defined noise regions. For each asset $j$ on day $t$, the combined upper band is defined as:
\begin{equation}
\text{UpperBand}_{t,j} = \min(\text{DonchianUp}_{t,j}, \text{KeltnerUp}_{t,j}),
\end{equation}
where:
\begin{align}
\text{KeltnerUp}_{t,j} &= \text{EMA}(P, n) + 1.4 \cdot k \cdot \frac{1}{n} \sum_{i=0}^{n-1} |\Delta P_{t-i,j}|, \\
\text{DonchianUp}_{t,j} &= \max(P_{t,j}, \ldots, P_{t-n+1,j}).
\end{align}
A long position is initiated when the closing price satisfies:
\begin{equation}
P_{t,j} \geq \text{UpperBand}_{t-1,j}.
\end{equation}

\subsubsection{Position Sizing and Constraints}
The strategy employs a volatility scaling approach to ensure equal risk contribution from each asset. The weight for asset $j$ is determined as:
\begin{equation}
\omega_{j,t} = \frac{\text{TargetVol}}{\sigma_{j,t}},
\end{equation}
where $\sigma_{j,t}$ is the 14-day rolling volatility of asset $j$. Total portfolio exposure is constrained by a maximum leverage limit:
\begin{equation}
\sum_{j=1}^N \omega_{j,t} \leq 200\%.
\end{equation}
If the leverage constraint is violated, weights are rescaled proportionally:
\begin{equation}
\omega^*_{j,t} = \frac{\omega_{j,t}}{\text{Exposure}} \cdot 200\%.
\end{equation}

\subsubsection{Exit Criteria}
Positions are exited based on a trailing stop mechanism, defined by the lower band:
\begin{equation}
\text{LowerBand}_{t,j} = \max(\text{DonchianDown}_{t,j}, \text{KeltnerDown}_{t,j}),
\end{equation}
where:
\begin{align}
\text{KeltnerDown}_{t,j} &= \text{EMA}(P, n) - 1.4 \cdot k \cdot \frac{1}{n} \sum_{i=0}^{n-1} |\Delta P_{t-i,j}|, \\
\text{DonchianDown}_{t,j} &= \min(P_{t,j}, \ldots, P_{t-n+1,j}).
\end{align}
The trailing stop evolves conservatively:
\begin{equation}
\text{TrailingStop}_{t+1,j} = \max(\text{TrailingStop}_{t,j}, \text{LowerBand}_{t,j}).
\end{equation}
A position is closed if:
\begin{equation}
P_{t,j} \leq \text{TrailingStop}_{t,j}.
\end{equation}

\subsection{Areas for Improvement}
The original strategy allocates uninvested cash to 1-month Treasury bills, providing a risk-free return when positions are not held. Notably, the initial strategy had 4276 such days since 2000 (47\% of the time), where cash was allocated to T-bills. While this mechanism effectively preserves capital and maintains productivity for idle funds, it may overlook opportunities to enhance returns through alternative cash management approaches. Additionally, several other aspects of the strategy could benefit from refinement. For example, its sensitivity to parameter choices, such as $n$, $k$, and $\text{TargetVol}$, highlights the challenge of balancing responsiveness to trends with robustness to noise. These factors, combined with the reliance on simple T-bill allocation, motivate the exploration of alternative methods aimed at improving risk-adjusted returns. This work forms part of a broader, iterative effort to address these areas and enhance the strategy’s adaptability to dynamic market conditions.

\section{Suggested Modifications}

\subsection{Fallback Strategy}
To address the challenge of cash allocation during no-investment days, we modified the original strategy to ensure that all capital remains invested at all times with no cash invested into T-bills. In the original strategy, signals failed to identify valid positions on 50 days since 2000, leading to uninvested cash and missed opportunities. To overcome this limitation, we implemented and evaluated three distinct fallback strategies: the moving average-based method, the risk-parity-based method, and a simple equal-weight allocation as a baseline for comparison. The strategy reassesses daily which assets to allocate capital to, ensuring that capital remains consistently deployed, even on days when primary signals fail.

The moving average-based fallback method uses short-term and long-term moving averages of asset returns to identify assets with the highest short-to-long-term ratios for investment. On days where this method still allocates all-zero weights, a second-tier fallback mechanism applies equal-weight allocation across all industry portfolios. The analysis varied the moving average window between 20 and 100 days, UP\_DAY thresholds between 5 and 20, and DOWN\_DAY thresholds between 15 and 40. Results are presented in Table~\ref{tab:moving_avg_fixed_final}, and implementation details can be found in \cite{mia2}.

The risk-parity-based fallback method allocates capital such that each asset contributes equally to the portfolio's overall risk on all-zero days. Weights are derived by inverting and normalizing asset volatilities. When volatility data is unavailable, a secondary fallback to equal-weight allocation ensures capital deployment. Results are shown in Table~\ref{tab:risk_parity}, and the implementation is detailed in \cite{mia3}.

For comparison, a pure equal-weighted fallback strategy was implemented as a baseline to assess the value added by the tiered fallback approaches. Results are summarized in Table~\ref{tab:equal_weight}, with implementation details available in \cite{mia4}.

\subsection{Added Momentum Signal}
As the results from the fallback layering proved insufficient, we restructured the strategy to incorporate a simple momentum signal alongside Donchian and Keltner bands. The momentum signal is based on the average returns of an industry portfolio over a rolling window of a specified length (MOMENTUM\_WINDOW), identifying the top 10 industries with the highest momentum. This ranking reduces reliance on fallback mechanisms by minimizing the occurrence of all-zero signal days to just 2 since 2000, compared to 50 in the original strategy. On these 2 days, a simple equal-weighted fallback mechanism was applied for capital deployment.

In this modified strategy, the weight of each industry portfolio is first determined by the momentum signal. If an industry is included, its weight is further adjusted proportionally based on inverse volatility, ensuring all capital remains invested. Results from this improved momentum strategy are presented in Table~\ref{tab:momentum_strategy_full}, and implementation details can be found in \cite{mia1}.

\subsection{Optimization of Rolling Periods and Volatility Target Parameter}
To enhance strategy performance, we optimized both the rolling periods for trend channels (e.g., Donchian or Keltner bands) and the target volatility parameter. Short rolling periods improve sensitivity to price movements, enabling faster trend detection but increasing turnover, noise, and transaction costs. In contrast, longer rolling periods smooth out market fluctuations, reduce costs, and emphasize long-term trends, albeit at the risk of delayed responses.

The 	extbf{volatility target parameter} adjusts position sizes to control risk exposure. Lower targets stabilize returns and limit drawdowns but may underutilize leverage, whereas higher targets amplify returns and risk, particularly in trending markets. Using an in-sample dataset, we optimized these parameters to maximize performance metrics such as the Sharpe Ratio and IRR. Results are illustrated in Figure~\ref{fig:Picture1}, and the implementation is detailed in \cite{sahaphon}.

\subsection{Equal-Weight Allocation}
Recognizing limitations in the original risk-parity model, which balances risk contributions but disregards expected returns, we transitioned to an equal-weight allocation. Risk-parity often overweights low-volatility assets that deliver lower returns and relies heavily on accurate volatility and correlation estimates, which fluctuate over time and can result in misallocations. Additionally, risk-parity may leave capital unallocated when high-volatility industries dominate the portfolio.

The 	extbf{equal-weight allocation} approach addresses these issues by uniformly allocating capital across industries, reducing reliance on unstable volatility estimates and mitigating concentration in low-volatility assets. Results are shown in Figure~\ref{fig:Picture2}, with implementation details provided in \cite{sahaphon}.

\subsection{S\&P 500 Allocation}
Given that equal-weight allocation did not consistently outperform risk-parity, we sought to address a key limitation of risk-parity: the presence of unallocated capital or "leftover money." Instead of investing this capital in low-yield T-bills, we allocated it to the S\&P 500 index to maintain market exposure, capture the long-term equity risk premium, and avoid opportunity costs.

The S\&P 500 allocation ensures diversification, positions the portfolio for market recoveries, and reduces reliance on precise timing decisions. To implement this, we dynamically tracked the entry and exit of stocks from the S\&P 500 index using data from the CRSP database (dsp500list). This eliminated survivorship bias by calculating returns based on stocks included in the index at any point in time. Daily price and return data were sourced from CRSP's daily stock file (dsf) and adjusted for corporate actions (e.g., splits and dividends). The index return was calculated using market-cap weights, ensuring robust and accurate performance representation. Results are displayed in Figure~\ref{fig:Picture3}, with full implementation details in \cite{sahaphon}.

\subsection{Industry Exclusion}
To improve portfolio performance, we introduced an industry exclusion process to eliminate underperforming sectors. Industry performance was assessed in the in-sample dataset using metrics such as Sharpe Ratio, IRR, hit ratio, maximum drawdown, and alpha. These metrics were standardized and combined into a composite score to rank industries.

Portfolios with varying numbers of excluded industries were constructed using in-sample validation data to identify the optimal composition. Rankings were derived from prior-period results, acknowledging that future performance cannot be predicted ex-ante. Results are shown in Figure~\ref{fig:Picture4} and Figure~\ref{fig:Picture5}, with implementation details provided in \cite{sahaphon}.

\subsection{Walk-Forward Analysis}
To evaluate the strategy's robustness and generalizability, we applied walk-forward analysis (WFA). This Point-in-Time approach prevents lookahead bias, reduces overfitting, and simulates real-world trading conditions. WFA divides the dataset into overlapping training and testing windows:
\begin{itemize}
    \item \textbf{Training Period}: 5 years of historical data used for parameter optimization.
    \item \textbf{Testing Period}: 1 year following the training period for out-of-sample (OOS) validation.
    \item \textbf{Iterative Shifting}: Windows are shifted forward by one year per iteration to ensure strict OOS separation.
\end{itemize}

Optimization employs a grid search over predefined hyperparameters, including UP\_DAY, DOWN\_DAY, ADR\_VOL\_ADJ, KELT\_MULT, and target volatility (target\_vol). Fixed parameters include initial capital $AUM_0 = 1$, maximum leverage of 2.0, and optional cash flow reinvestment. The Sharpe Ratio (SR) is the primary metric for optimization, calculated as:
\begin{equation}
SR = \frac{\mathbb{E}[r_t - r_f]}{\sigma(r_t - r_f)} \sqrt{252},
\end{equation}
where $r_t$ is portfolio return, $r_f$ is the risk-free rate, and $\sigma$ is the standard deviation of excess returns.

At each iteration, the best-performing parameter set from the training period is applied to the subsequent OOS period. Results across iterations are aggregated to evaluate performance stability. To enhance computational efficiency, optimizations are parallelized across multiple CPU cores.

\begin{figure}[H]
\centering
\includegraphics[width=.6\textwidth]{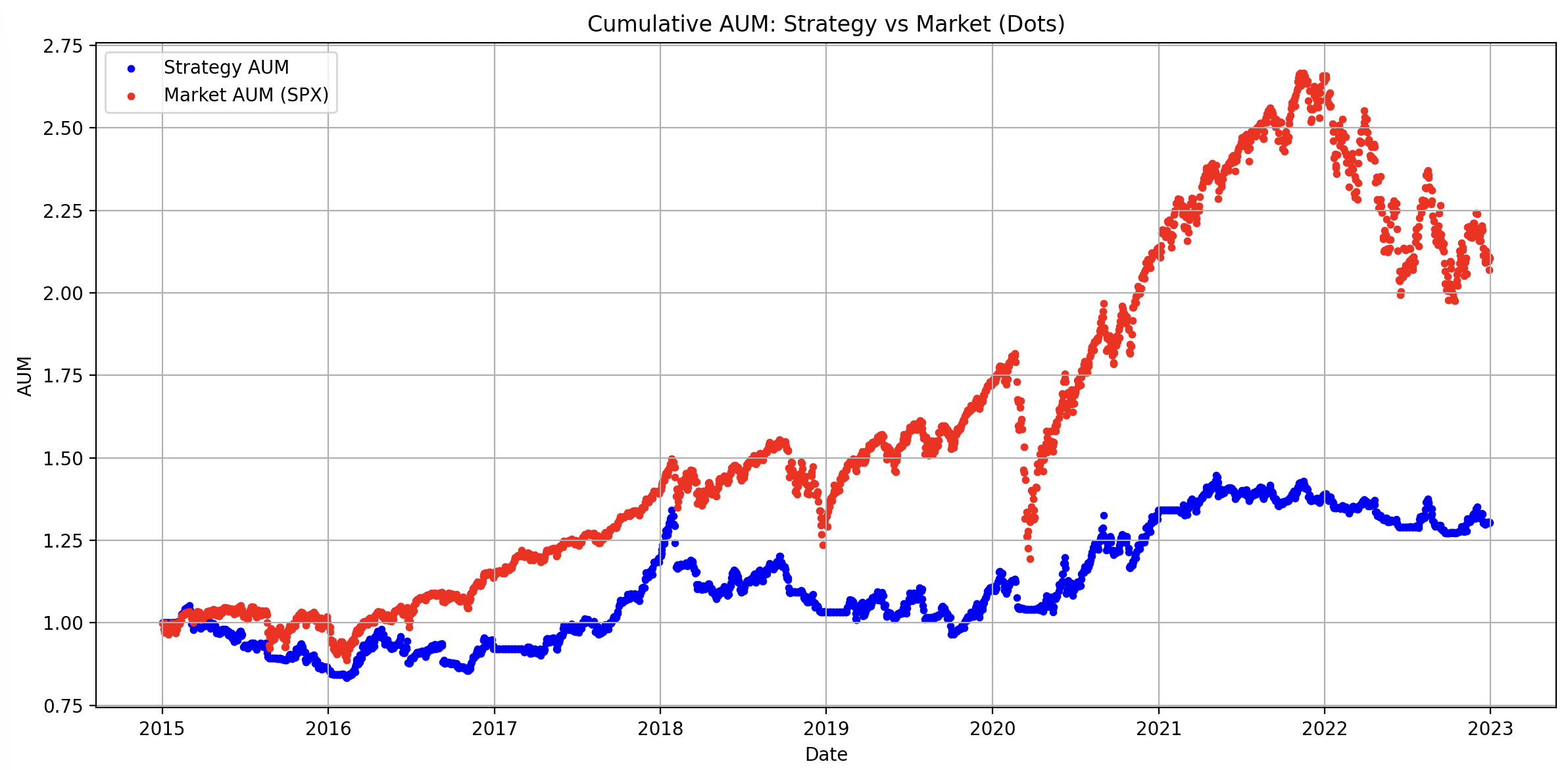}
\caption{Performance of the Original paper's startegy compared to the market, when run on WFA}
\label{fig:wfa_results}
\end{figure}

WFA results are presented in Figure~\ref{fig:wfa_results}. Full implementation details are available in \cite{github_repo}.

\section{Results}

Results from the Moving Average-Based Fallback method are displayed in Table~\ref{tab:moving_avg_fixed_final}.

\begin{longtable}{ll|cc|cccc}
\caption{\textbf{Method 1 -- Moving Average-Based Fallback}}
\label{tab:moving_avg_fixed_final} \\
\toprule
\multicolumn{2}{l|}{\textbf{UP\_DAY}} & \multicolumn{6}{c}{5} \\ 
\multicolumn{2}{l|}{\textbf{DOWN\_DAY}} & \multicolumn{6}{c}{15} \\
\midrule
\textbf{ma\_window} & & \multicolumn{2}{c|}{20} & \multicolumn{4}{c}{100} \\ 
\textbf{Data} & & is\_data & isv\_data & is\_data & isv\_data & v\_data & pt1\_data \\
\midrule
Alpha & & 2.05 & \cellcolor{red!15}-3.23 & 1.57 & 1.72 & 2.73 & \cellcolor{red!15}-6.91 \\
Beta & & 0.84 & 0.89 & 0.83 & 0.71 & 0.8 & 0.87 \\
Worst Day & & -28.92 & -16.79 & -11.07 & -16.79 & -4.13 & -2.17 \\
Best Day & & 41.18 & 10.87 & 9.7 & 11.99 & 4.64 & 2.35 \\
\bottomrule
\end{longtable}

\begin{longtable}{ll|c|cccc}
\toprule
\multicolumn{2}{l|}{\textbf{UP\_DAY}} & \multicolumn{5}{c}{20} \\ 
\multicolumn{2}{l|}{\textbf{DOWN\_DAY}} & \multicolumn{5}{c}{40} \\
\midrule
\textbf{ma\_window} & & \multicolumn{1}{c|}{20} & \multicolumn{4}{c}{100} \\ 
\textbf{Data} & & is\_data & is\_data & isv\_data & v\_data & pt1\_data \\
\midrule
Alpha & & \cellcolor{red!15}-0.33 & 2.92 & 12.35 & 4.54 & \cellcolor{red!15}-9.9 \\
Beta & & 0.79 & 0.76 & 0.55 & 0.76 & 0.87 \\
Worst Day & & -28.92 & -10.9 & -9.07 & -6.39 & -2.46 \\
Best Day & & 10.59 & 11.32 & 11.99 & 5.55 & 2.36 \\
\bottomrule
\end{longtable}

Results from the risk-parity-based fallback method are displayed in Table~\ref{tab:risk_parity}.

\begin{longtable}{lcccc|cc}
\caption{\textbf{Method 2 -- Risk-Parity-Based Fallback}}
\label{tab:risk_parity} \\
\toprule
\textbf{Parameter}  & \multicolumn{4}{c|}{UP\_DAY = 20, DOWN\_DAY = 40} & \multicolumn{2}{c}{UP\_DAY = 5, DOWN\_DAY = 15} \\
\midrule
Data                & is\_data & isv\_data & v\_data & pt1\_data & is\_data & isv\_data \\
\midrule
Alpha               & 1.84     & 2.65      & 2.77    & \cellcolor{red!15}-9.9 & 1.34     & \cellcolor{red!15}-5.93 \\
Beta                & 0.8      & 0.84      & 0.76    & 0.87      & 0.83     & 0.89 \\
Worst Day           & -9.42    & -11.44    & -6.39   & -2.46     & -11.07   & -16.79 \\
Best Day            & 10.87    & 10.41     & 5.55    & 2.36      & 10.87    & 7.25 \\
\bottomrule

\end{longtable}

Table~\ref{tab:equal_weight} summarizes the findings from the equal weight fallback.

\begin{longtable}{lcccc|cc}
\caption{\textbf{Method 3 -- Equal Weight Fallback}}
\label{tab:equal_weight} \\
\toprule
\textbf{Parameter}  & \multicolumn{4}{c|}{UP\_DAY = 20, DOWN\_DAY = 40} & \multicolumn{2}{c}{UP\_DAY = 5, DOWN\_DAY = 15} \\
\midrule
Data                & is\_data & isv\_data & v\_data & pt1\_data & is\_data & isv\_data \\
\midrule
Alpha               & 1.66     & 2.36      & 3.58    & \cellcolor{red!15}-10.19 & 1.62     & \cellcolor{red!15}-5.85 \\
Beta                & 0.8      & 0.84      & 0.76    & 0.88      & 0.84     & 0.89 \\
Worst Day           & -9.42    & -11.44    & -6.39   & -2.79     & -11.07   & -16.79 \\
Best Day            & 10.87    & 10.41     & 5.55    & 2.36      & 10.87    & 7.25 \\
\bottomrule
\end{longtable}

As shown in Table \ref{tab:momentum_strategy_full}, MOMENTUM\_WINDOW (MW) was modified to find the best strategy.

\scriptsize
\renewcommand{\arraystretch}{1.2}
\begin{longtable}{p{1.2cm}|p{1.1cm}p{1.1cm}|p{1.1cm}p{1.1cm}p{1.1cm}p{1.1cm}|p{1.1cm}p{1.1cm}p{1.1cm}p{1.1cm}}
\caption{\textbf{Analysis of Modified Strategy with Momentum}}
\label{tab:momentum_strategy_full} \\
\toprule
\textbf{MW} & \multicolumn{2}{c|}{\cellcolor{yellow!30}\textbf{20 days}} 
            & \multicolumn{4}{c|}{\cellcolor{purple!10}\textbf{60 days}} 
            & \multicolumn{4}{c}{\cellcolor{blue!10}\textbf{360 days}} \\
\midrule
\textbf{TIME} & \textbf{is\_data} & \textbf{isv\_data} 
              & \textbf{is\_data} & \textbf{isv\_data} & \textbf{v\_data} & \textbf{pt1\_data} 
              & \textbf{is\_data} & \textbf{isv\_data} & \textbf{v\_data} & \textbf{pt1\_data} \\
\midrule
\textbf{Alpha} & \cellcolor{green!20}2.88 & \cellcolor{red!15}-0.91 
               & \cellcolor{green!20}3.04 & \cellcolor{green!20}0.53 & \cellcolor{green!20}7.82 & \cellcolor{red!15}-8.19 
               & \cellcolor{green!20}3.7 & \cellcolor{green!20}1.54 & \cellcolor{green!20}6.0 & \cellcolor{red!15}-6.39 \\
\midrule
\textbf{Beta}  & 0.8 & 0.84 
               & 0.81 & 0.87 & 0.78 & 0.91 
               & 0.84 & 0.9 & 0.79 & 0.93 \\
\midrule
\textbf{Worst Returns} & -8.77 & -8.13 
                          & -8.91 & -9.53 & -4.06 & -2.18 
                          & -8.93 & -8.84 & -4.44 & -2.18 \\
\midrule
\textbf{Best Returns} & 9.46 & 7.68 
                         & 7.77 & 8.38 & 4.67 & 2.57 
                         & 12.81 & 8.44 & 4.67 & 2.54 \\
\bottomrule
\end{longtable}

\begin{figure}[H]
\centering
\includegraphics[width=1\textwidth]{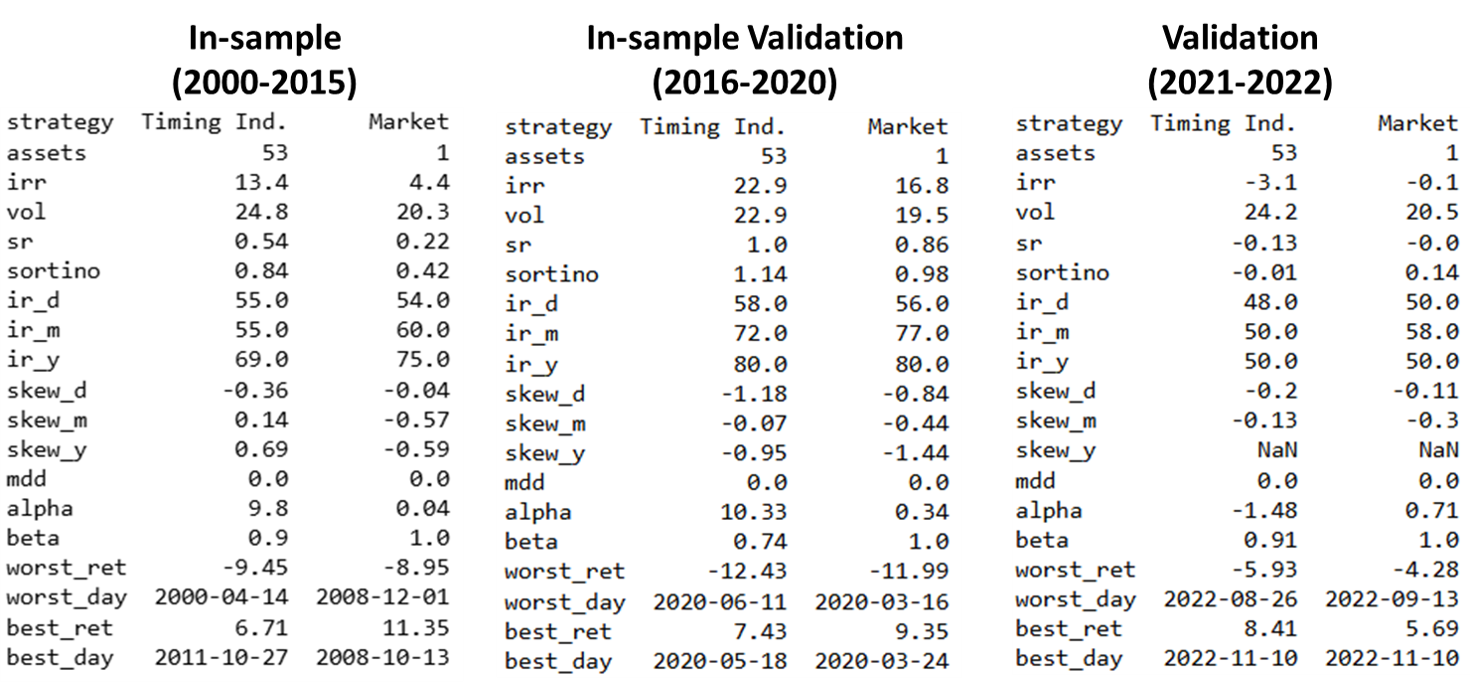}
\caption{Performance metrics for optimizing rolling periods and volatility target parameters across test sets}
\label{fig:Picture1}
\end{figure}

\begin{figure}[H]
\centering
\includegraphics[width=1\textwidth]{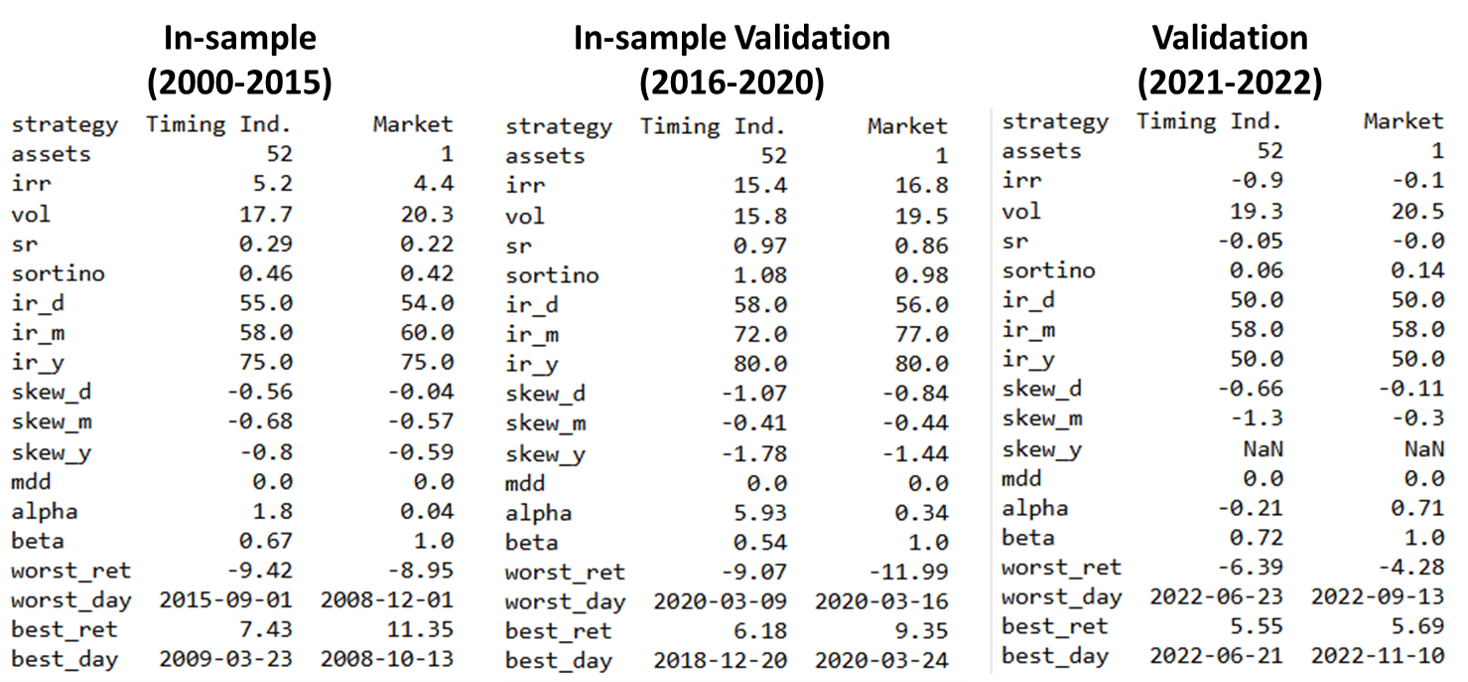}
\caption{Performance metrics for Equal-Weight Allocation across test sets}
\label{fig:Picture2}
\end{figure}

\begin{figure}[H]
\centering
\includegraphics[width=.6\textwidth]{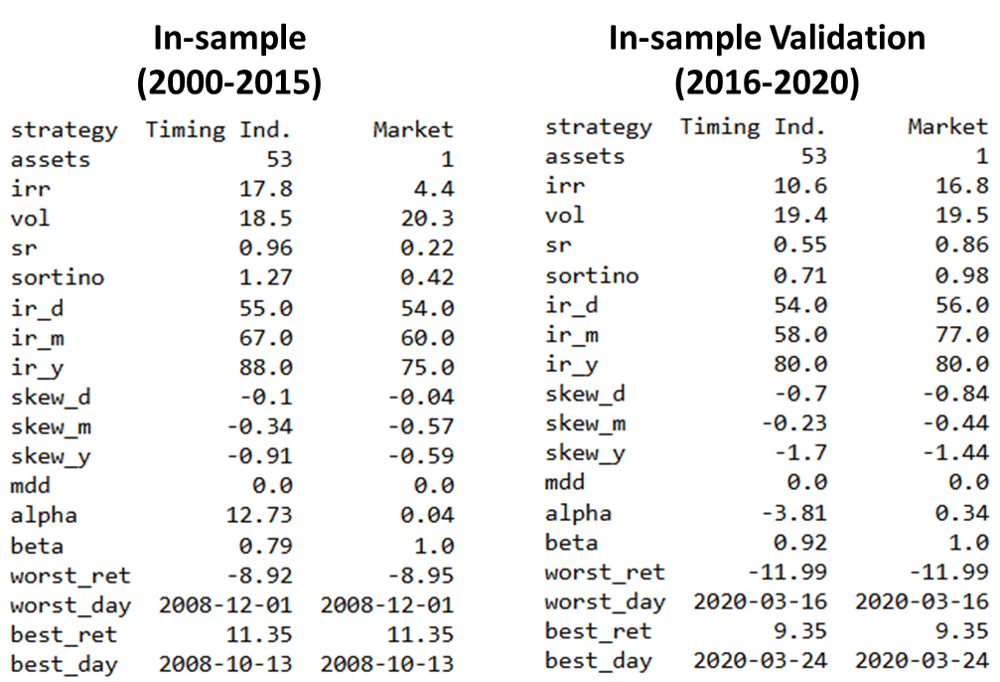}
\caption{Performance metrics for S\&P 500 Allocation across test sets}
\label{fig:Picture3}
\end{figure}

\begin{figure}[H]
\centering
\includegraphics[width=.6\textwidth]{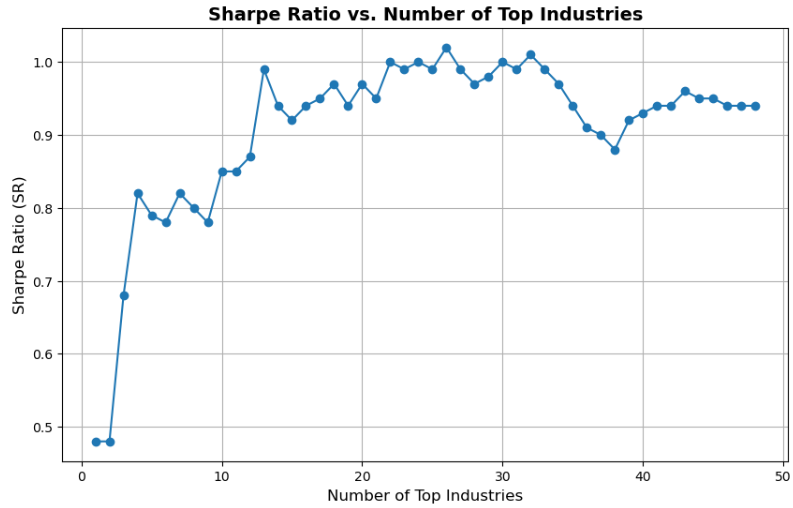}
\caption{Sharpe ratio across varying numbers of top-k industries included in the strategy}
\label{fig:Picture4}
\end{figure}

\begin{figure}[H]
\centering
\includegraphics[width=.6\textwidth]{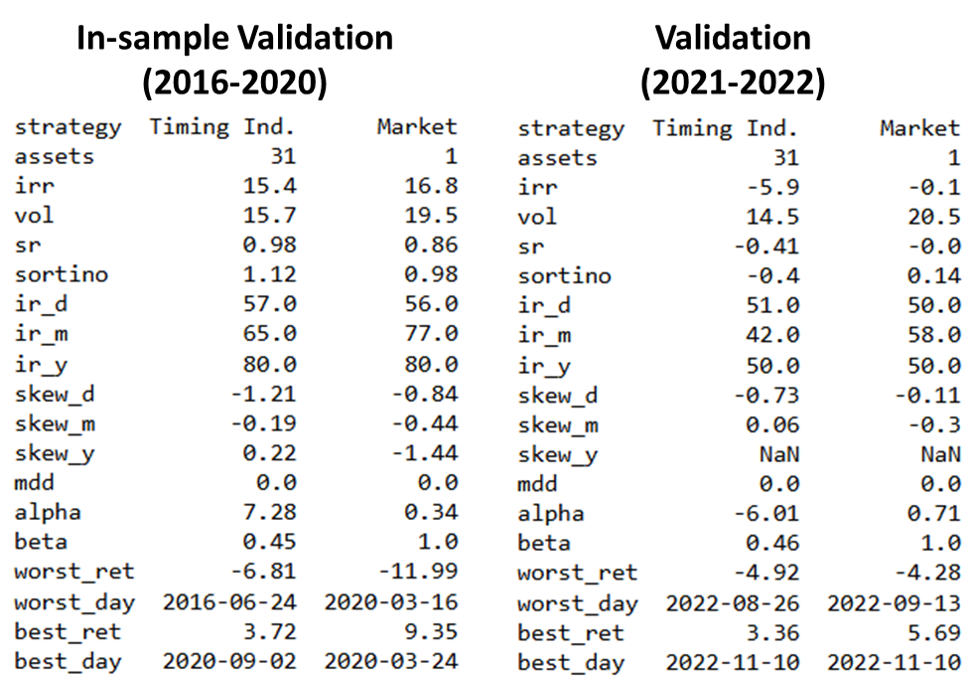}
\caption{Performance metrics for Industry Exclusion across test sets}
\label{fig:Picture5}
\end{figure}

\normalsize

\section{Discussion}

\subsection{Fallback Strategy Analysis}
The fallback strategy modifications aimed to address the capital allocation challenge during no-investment days. Despite extensive testing, none of the attempted combinations produced satisfactory results beyond \textit{v\_data}. While volatility windows were tested at higher values, their impact on outcomes was negligible. Consequently, we focused on presenting tests with varying UP\_DAY and DOWN\_DAY thresholds. Among these, the moving-average-based fallback method emerged as the most effective, with performance comparable or superior to the risk-parity-based fallback (Table~\ref{tab:moving_avg_fixed_final}). This highlights the method's relative simplicity and robustness as the best fallback option.

\subsection{Momentum Strategy Analysis}
The integration of a momentum signal significantly improved returns compared to the fallback strategies. However, the enhancements were insufficient to sustain performance beyond the initial paper trading period. Despite promising in-sample and validation results, particularly for longer momentum windows of 60 and 360 days, the strategy's performance deteriorated during live trading. This underperformance underscores the limitations of the original framework and the challenge of developing momentum signals that generalize across market conditions.

\subsection{Rolling Periods and Volatility Target Analysis}
The optimization of rolling periods and volatility targets delivered strong results in the in-sample and initial validation phases. However, these parameters failed to generalize when tested on separate validation datasets, resulting in weaker Sharpe ratios and IRRs (Figure~\ref{fig:Picture1}). This outcome suggests overfitting to the in-sample data and emphasizes the difficulty of achieving robust performance across dynamic markets. The iterative exploration of alternative optimization approaches highlights the need for further refinement to improve generalizability.

\subsection{Equal-Weight Allocation Performance}
Equal-weight allocation served as a simplified alternative to the risk-parity method, performing competitively in the in-sample dataset and initial validation phases. However, this approach underperformed in the separate validation dataset, reinforcing the challenges of achieving consistency across diverse market environments. This performance discrepancy indicates that equal-weight allocation, while straightforward, may not provide the robustness needed for dynamic trend-following strategies (Figure~\ref{fig:Picture2}).

\subsection{S\&P 500 Allocation Performance}
The S\&P 500 allocation strategy sought to address the issue of unallocated capital by maintaining market exposure. While it delivered strong results in the in-sample dataset, its performance weakened during in-sample validation, indicating potential overfitting and a lack of generalizability. These findings highlight the need for further adjustments to improve the robustness of this approach and ensure consistent performance in out-of-sample scenarios (Figure~\ref{fig:Picture3}).

\subsection{Industry Exclusion Performance}
The introduction of industry exclusion aimed to eliminate underperforming sectors and optimize portfolio composition. While the approach yielded promising results in the in-sample dataset, it failed to generalize to the separate validation dataset, with notable underperformance. This pattern suggests overfitting to historical data and highlights the limitations of industry exclusion strategies in predicting future performance. Despite initial success, further refinements are required to enhance robustness and adaptability (Figures~\ref{fig:Picture4} and \ref{fig:Picture5}).

\subsection{Walk-Forward Analysis}
The Walk-Forward Analysis (WFA) provides a realistic assessment of the strategy's performance under strict out-of-sample conditions. While the strategy demonstrated the ability to preserve capital during specific periods, its cumulative performance fell short of market benchmarks, as shown in Figure~\ref{fig:wfa_results}. 

The strategy's underperformance can be attributed to rising sector correlations, increasing market efficiency, and the limited effectiveness of historical industry momentum signals in modern environments. Conservative volatility targeting and fallback mechanisms, designed to reduce risk, may have further constrained upside potential during strong upward trends. 

Overall, the WFA results underscore the challenges of adapting historical trend-following strategies to dynamic market conditions. These findings represent a critical step in an iterative research process, highlighting areas for future enhancements, such as alternative risk-adjusted allocation schemes, improved parameter tuning, and refined momentum indicators, to better capture evolving market trends.

\section{Summary}
This study revisited and modified the long-only industry trend-following strategy proposed by Zarattini and Antonacci. The modifications addressed key limitations, including reliance on T-bills, the occurrence of all-zero signal days, and overfitting to historical data. Several alternative strategies were explored:

\begin{itemize}
    \item The \textbf{moving-average-based fallback} emerged as the most effective mechanism for maintaining capital allocation.
    \item A \textbf{momentum signal} reduced reliance on fallback methods but still faced challenges in generalizing across live trading periods.
    \item Optimization of \textbf{rolling periods and volatility targets} demonstrated the difficulty of balancing sensitivity and stability without overfitting.
    \item \textbf{Equal-weight allocation} and \textbf{S\&P 500 fallback} approaches improved capital deployment but lacked robustness in out-of-sample validation.
    \item \textbf{Industry exclusion} provided performance enhancements in-sample but failed to generalize effectively.
    \item \textbf{Walk-Forward Analysis} underscored the strategy's limitations in capturing market trends under evolving conditions.
\end{itemize}

While the proposed enhancements demonstrated incremental improvements, the results highlight persistent challenges in adapting historical trend-following methods to modern markets. Rising sector correlations, increasing market efficiency, and overfitting remain significant obstacles. Future research should focus on refining risk-adjusted allocation strategies, enhancing parameter tuning, and developing robust momentum signals to improve generalizability and performance.

\section{Acknowledgements}
We extend our heartfelt gratitude to our professor, Naftali Cohen, whose direct and pragmatic approach has profoundly influenced our perspective on academic readings, encouraging a clear and focused methodology. 
We also extend our sincere appreciation to Columbia University for providing us with the resources, facilities, and supportive environment that made this research possible. 

\nocite{*}
\bibliography{references}

\end{document}